# Unicorns Do Not Exist: Employing and Appreciating Community Managers in Open Source

**Raphael Sonabend[1, 2], Anna Carnegie[3], Anne Lee Steele[4], Marie Nugent[5], Malvika Sharan[4, 6]**

**Raphael Sonabend[1, 2]**

raphaelsonabend@gmail.com

0000-0001-9225-4654

**Anna Carnegie[3]**

anna.carnegie@kcl.ac.uk

0000-0002-6385-7795

**Anne Lee Steele[4]**

asteele@turing.ac.uk

0000-0002-6902-7768

**Marie Nugent[5]**

marie.nugent@genomicsengland.co.uk

0000-0001-7962-6679

**Malvika Sharan[4, 6]**

msharan@turing.ac.uk

0000-0001-6619-7369

1. OSPO Now, W1T 3EY, United Kingdom.

2. Imperial College London, Exhibition Rd, South Kensington, London, SW7 2AZ, United Kingdom.

3. Social, Genetic and Developmental Psychiatry Centre, King's College London, Memory Lane, London SE5 8AF, United Kingdom.

4. Alan Turing Institute, 96 Euston Road, London NW1 2DB, United Kingdom.

5. Genomics England, 1 Canada Square, London E14 5AB, United Kingdom.

6. Open Life Science (OLS), 31 Eaton State, Wimblington, PE15 0QE March, United Kingdom.

**Abstract**

Open-source software is released under an open-source licence, which means the software can be shared, adapted, and reshared without prejudice. In the context of open-source software, community managers manage the communities that contribute to the development and upkeep of open-source tools. Despite playing a crucial role in maintaining open-source software, community managers are often overlooked. In this paper we look at why this happens and the troubling future we are heading towards if this trend continues. Namely if community managers are driven to focus on corporate needs and become conflicted with the communities they are meant to be managing. We suggest methods to overcome this by stressing the need for the specialisation of roles and by advocating for transparent metrics that highlight the real work of the community manager. Following these guidelines can allow this vital role to be treated with the transparency and respect that it deserves, alongside more traditional roles including software developers and engineers.

**Statements and Declarations**

There are no competing interests. The views expressed in this paper are those of the authors and not necessarily those of their employers or funders.

**Funding acknowledgement**

MS and ALS' role has been supported by the Ecosystem Leadership Award under the EPSRC Grant EP/X03870X/1 & The Alan Turing Institute. The funders had no role in the preparation or decision to publish the manuscript.

## Introduction

It's possible, and not altogether unreasonable, to imagine a world in the not-too-distant future, where in place of kind, caring, and engaged community managers, there are instead "customer service call centres" for open source community management. Except, instead of infinite rows of people on the phone, community managers are messaging on Slack, running webinars, and drinking pints of virtual coffee. This dystopian future might sound like an impossibility for a community that would think of itself as altruistic (Baytiyeh and Pfaffman, 2010), but in light of similarly mechanised services such as UpWork and Mechanical Turk, this trajectory is not so impossible to imagine for the wider open source world. In this paper, we will outline why we think this is the case and how it may be prevented.

Academia and industry have seen an increased uptake of open-source solutions, which have a myriad of benefits for both sectors including saving on development time (by re-using existing software) and costs (as proprietary licences are often expensive) (Chesbrough et al., 2023). On the other hand, open source can also present risks, including legal complexities around licensing (Brock, 2022), software dependencies, maintaining code over time (Eghbal, 2020), and the need for bespoke sustainable solutions to mitigate these challenges (Tamburri et al., 2019).

These are real, existential risks for any open-source project that has to be considered when developing (or sometimes even just using) open-source software. "Abandonware" occurs when a piece of software is developed and then left unmaintained. Eventually, bugs are found and the software becomes unusable for new users. Any 'downstream' software can then start to break, creating disastrous knock-on consequences as the number of downstream dependencies increases. Abandonware is increasingly common as software is more routinely created as a part of scientific research. UC San Diego estimated that faculty and students have contributed to 32,000 public repositories; only a small fraction of which need to be influential for these to pose risks when said researchers leave (if they haven't already). As well as mitigating upstream risks, organisations also need to constantly identify and fix bugs in the code they develop or use whilst simultaneously ensuring new features are being added to accommodate client needs. Instead of relying on researchers and software engineers to find and fix bugs in abandonware and their codebase, organisations can instead depend on communities. For example, bug bounty programmes provide a cost-efficient method for companies to pay people one-off fees for detecting security flaws in their code. In an increasingly automated and online world, consumers are generally happy to report bugs and suggest features for no incentive beyond the satisfaction that they are improving a product. Open-source software may be cheaper to produce, but it can also be harder to generate revenue from. Therefore, long-term sustainable solutions are usually required to ensure open source code can be maintained over time. In academia, funding bodies are often reluctant to provide 'maintenance funding', instead preferring research institutions to generate innovative technical solutions as well as resources for maintaining them in the long term.

An increasingly common method for organisations to tackle these risks is to depend on a community of volunteers led by a community manager employed by the organisation (Michlmayr, 2009). There is good evidence that building a community can be a low-cost, ethical and practical method of sustaining open-source software (Curto-Millet and Jiménez, 2023). However, the relative speed at which communities of volunteers are growing is coming at the expense of the people who lead them. Whether that's overlooking the value of maintenance and 'caregiving' (Russell and Lee, 2019), a range of skills that community managers are expected to bring to 'do it all', or job descriptions that confuse research, brand, and even crypto community managers; all of this leads to overwhelmed and underappreciated community managers.

## What actually is the role of a community manager in open source?
The definition of a community manager differs across sectors but within open source projects, at the most basic level, a community manager is someone who: i) builds a community around an open source product; ii) keeps the community engaged and excited via accessible events such as webinars, virtual meet-ups, and more; iii) celebrates successes in the community to create an empowered group of people. In short, a community manager for an open-source project is responsible for making sure the 10s, 100s, or 1000s of people who *want* to use and contribute to your code, can easily do so whilst feeling motivated, rewarded and engaged.

This is a vital job role and requires someone with good interpersonal skills, passion for the project, vision for how they see the community developing, an ability to engage different stakeholders at different levels of technical proficiency, and numerous other technical, interpersonal and communication skills (Woodley et al., 2021). A community manager for an open-source project must be able to speak with developers, users, and beneficiaries of software, who will range from having no coding experience to using extensive technical jargon.

## The misunderstood and unseen labour of community managers

Despite the extensive criteria required to succeed in their roles, community managers' overall contributions to the organisation's success can remain unseen (D'Ignazio and Klein, 2020). In addition to the labour that goes into creating tangible outputs, the community manager role also involves significant emotional labour that may be overlooked by others (D'Ignazio and Klein, 2020). As convenors of safe spaces for volunteers, the role often lends itself to becoming a 'friendly ear' for community members to confide in and seek support from. The emotional and social support provided by community managers is rarely seen by their employers who are often at a distance from the community itself and to compound matters, community managers are rarely provided with tailored mental health support or training.

Open source is also impacted by the social inequities, lack of diversity and gender divides that typically affect the STEM (science, technology, engineering and mathematics) workforce (Chełkowski et al., 2016; Gasparini et al., 2020). Although long analysed and extensively critiqued, professional roles in engineering or computer science that are deemed highly "technical" are categorised differently from roles that are predominantly attributed with "transferable" skills, often labelled as "soft" skills, such as those required for community managers. Such hierarchical classifications of professions are both cause and consequence of occupational segregation in male-dominating industries, where roles stereotypically associated with maintenance, care and relationship-building, like community management, are predominantly held by women and minority groups (Mintz and Krymkowski, 2010; Reskin and Cassirer, 1996; Campero, 2021; Szlavi et al., 2023). This is most likely the reason why community manager roles are perceived as lower in status and less well-paid compared to software developer roles.

Contrary to the perception of soft skills, there is an inherent need to apply highly complex technical and social skills to enhance belonging, intersectionality, and inclusivity within open source teams and communities. This directly contributes to broadening participation and embedding human-centric considerations in the development and use of technology, practices and infrastructure (Carter et al., 2021). Hence, roles such as community managers are crucial for cultivating inclusive communities and creating opportunities for diverse members, including those from minority groups, to participate in and integrate a broad range of perspectives in open source projects.

## How not to hire a community manager

To illustrate the problem we face today, let's take a step back and consider what the makeup of an open source team might look like. To keep this simple (and also to reflect academic and start-up sectors) we will consider a hypothetical research project focused on an open-source product. In this scenario, the most basic team requires two skills: domain expertise in the area of interest, and software development to write code and release it publicly (such as on GitLab) with an open-source licence. In practice, if a research project has received a substantial amount of funding (a single open-source project could receive millions from funders internationally, such as from the Chan Zuckerberg Initiative Essential Open Source Software funding call), then teams are likely to also have domain experts and developers. As funders, for example Wellcome and the UKRI, increasingly mandate that projects must be impactful and sustained over time, it is common to find a principal investigator (PI) who will be primarily tasked with demonstrating impact (usually in the form of the number of users or academic papers). The community manager will often be tasked with building an open source community with the goal of achieving the holy grail of self-sustaining open-source software.

As a result of this setup, community managers are often expected to also be skilled in coding (at least in the language used to develop the code), open source practices (including licensing, code readability and accessibility), strategic planning for long-term community development, and the core duties of building and engaging a community. To find this broad and incredibly diverse skills mix in *one* individual is next to impossible. Moreover, were the 'perfect candidate' to exist, it would in actuality be detrimental to the future of the project. Consider how much work needs to go into direct relationship-building and 'emotional care work', often termed 'softer' aspects of community engagement. Managers are usually expected to hold weekly (at the very least fortnightly) meetings with the community to keep them informed, 1:1 meetings with members if they have questions or concerns about the community or product, webinars and events to onboard and offboard new and old community members *and* activities to find new members such as by attending conferences, other community events or workshops. There is no way for a community manager to find time to do all of this whilst simultaneously ensuring that the code itself is accessible, community documentation is up to date, issues and pull requests on public repositories are being read, and that the interaction between in-house developers (if they have them) and the community is seamless, effective, and (this is often overlooked but is very important) polite.

To make matters worse, where metrics exist to assess the quality of a community manager's role, these tend to focus on quantitative results such as the number of events and attendees, and less so on the tone of meetings and greater community health; the latter often requiring a greater emotional effort on the community manager's part whilst also being largely invisible. Organisations such as CHAOSS are working towards building more inclusive metrics but until these are ubiquitous, community managers will struggle to demonstrate the true effort they put into their day-to-day role.

Finally, the amount of time required for carrying out community management tasks is often underestimated, contributing to either a high workload with multiple responsibilities or roles designed for part-time jobs. This is partly due to an under-appreciation of the role contributing to a lack of recognition as to why this should be a standalone full-time job, and partly because, as discussed above, the role is often taken on by members of minority groups, who in turn are more likely to work part-time (Catney and Sabater, 2015; Perez, 2019). This only compounds problems related to individuals not feeling valued, adding to job dissatisfaction and poor mental health, as the true responsibilities within the role, and the additional responsibilities outside the role of these professionals, remain invisible (Perez, 2019).

Underappreciating community managers doesn't just affect their well-being and job satisfaction, it can have serious repercussions for the community. If a community manager perceives that much of their work will go unseen and undervalued, their dedication can falter and loyalties can become split.

Let us play devil's advocate for one moment and very superficially depict how a purpose-built community for software can be viewed:

> "The purpose of building a community is to make use of free labour to sustain software over time by:
> 1. Making the community dependent on the software; and
> 2. Making the community members dependent on the community."

Phrased this way, open source communities are exploited to serve a larger organisation without any justification. So let's instead turn away from this interpretation and now consider a community as follows:

> "The purpose of building a community is to rely on crowdsourcing from volunteers to sustain software by:
>
> 1. Releasing free and impactful software that the community *want* to use and give back to by paying in time and not cash; and
> 2. Creating a sense of purpose and reward for contributing to the community."

If a community manager is viewed as a 'corporate stooge' whose primary loyalties lie to the organisation and not the community, then the first interpretation appears much more realistic. Whereas someone deeply embedded in the community who is genuinely concerned with, and passionately cares about community members, is more likely to be able to succeed in instilling the latter perspective. Not only does this matter for community health, it is crucial for the job satisfaction of the community manager. If you want to hire someone to maintain a community and to be honest and transparent, their loyalties must be with the community. If you want a community to be involved and volunteer their time in building something useful for them, then their perspective must come first. In other words, the customer (even when they are not paying) must always come first. It is therefore imperative that the community manager remains just that, a *community* manager, and that their responsibilities lie solely in advocating for the community's interests.

## Community managers all the way down

We return once again to our infinite row of community managers in a single service centre. Each is in charge of a different community with its own timetable of regular calls and meetups, webinars and virtual coffees, onboarding and offboarding sessions, and office hours for comments and complaints. Recognising the underappreciation of community managers makes this future much easier to imagine.

Consider an organisation that develops open-source software for a given area of research and is successful enough to have built a community, which is happily self-organising to sustain and maintain software. This community will still need *at least one* salaried individual to be the voice of the community, to ensure a sense of continuity, to merge bug fixes, and to work towards sustaining the community. This naturally falls to the project or the product lead who may have to wear the community manager hat. For bigger projects, multiple community managers might be required, and for large organisations releasing multiple open-source projects, even more. As the development of open-source products continues to grow, and their community managers continue to be ignored, what stops a large company from building a floor of, unenthused, emotionally drained, community managers sitting on Slack and Teams and Zoom, putting on fake smiles, and growing increasingly dissatisfied without anyone ever noticing?

This, we believe, is where we are. Organisations are searching for unicorns to manage *everything* community-related without caring about the community managers themselves. But it doesn't need to be this way: with specialisation and clearer objectives for what is required of the community and its manager(s), there is better hope for the long-term sustainability of the product, its community, and the managers that look after it.

## The Need for Specialisation

By definition, unicorns do not exist. Similarly, the skill set outlined above is hard to find, and as already stated, unlikely to be productive if not carefully delegated. The better solution is to find horses, narwhals, rhinos, and goats – a team of experts with different skill sets that can work well together. In other words, there is a need for specialisation of community manager roles and clarity of where they sit in research projects.

Before establishing specialisation within the community manager role we will first distinguish between community managers and software engineers or developers. Firstly, software developers/engineers tend to be hired to create new and novel tools and technologies with shorter-term objectives, for example developing infrastructure or tools to solve a real-world problem. People within software engineer/developer roles might have titles including *research software engineer*, *data scientist* (or wrangler, or steward), *machine learning engineer*, *research infrastructure developer*, and more. Typically these roles are hired to develop tools for one project and then reassigned once the product is developed. On the other hand, community manager-type roles are more likely to focus on developing, building, and maintaining a community, for the purpose of long-term sustainability and maintenance of the tools. Hence, software developers will typically be hired to have short-term outlooks to build tools, whereas community managers are hired with long-term outlooks to build communities. Both are therefore essential to long-term maintenance. Poorly built tools will not attract a community of users. This separation of roles creates a grey area, which can cause a lack of accountability in software development. Namely questions around where responsibility lies when it comes to end-user mapping to design user interfaces, maintenance of online repositories, and other areas at the point at which community members can act as developers. We believe this grey area should be filled by specialised community manager roles, who work closely alongside software developers and provide continuity even when the developers get reassigned to other projects. Therefore there is a needed gradient in terms of technical ability and areas of focus within software developers and community managers (see Figure 1).

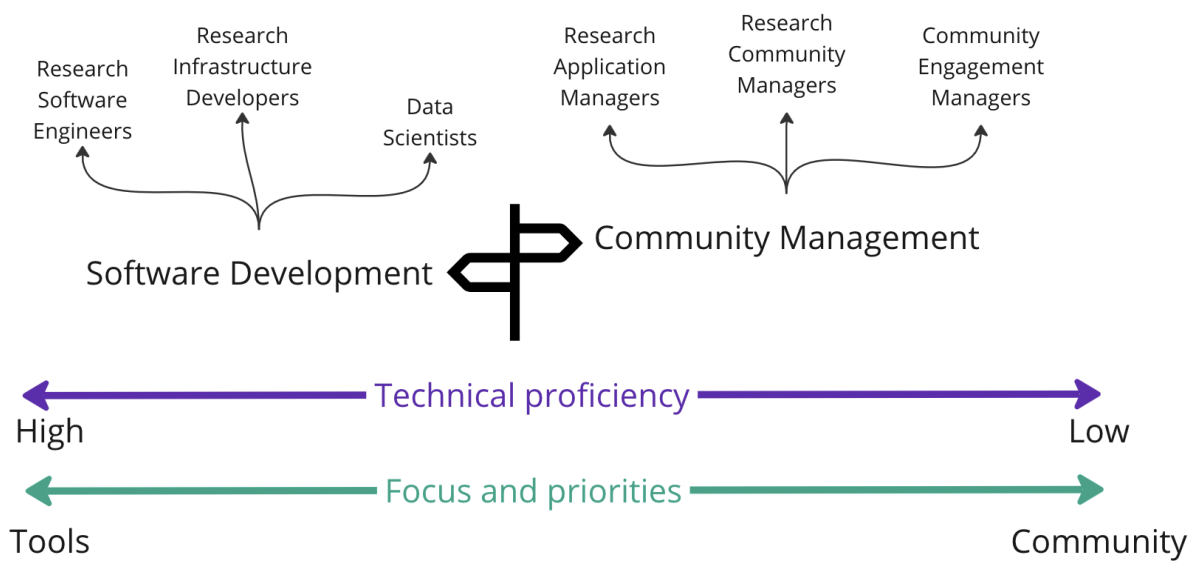

Figure 1: Representing specialised roles in software development. Moving from left to right roles decrease in their need to be technically proficient as their priorities shift from tools to community.

The proposed specialisation of community managers is partly based on the positions advocated in The Turing Way, these are: *Product Managers (PMs) or their equivalent Research Application Managers (RAMs)*, *Research Community Managers (RCMs)*, and *Community Engagement Managers (CEMs)*. PMs and RAMs, which may also be known as *Open Source Managers* (OSMs) or *Technical Community Managers* (TCMs), are focused on the intersection described above, responsibilities of these roles may include strategic decisions for community building and sustaining software and end-user mapping to iteratively improve and co-develop a tool's design. They work with other community manager roles to manage issue trackers and new pull requests, upskill users and facilitate the use of open source best practices including reproducibility, licensing, documentation, and more. These roles work closely with software developers to understand possibilities for tool development but their focus should be on community needs to avoid the "devil's advocate" position outlined above. CEMs are likely to have the least technical experience in the community manager roles but instead focus on ongoing engagement with the community and other partners. The responsibilities of the CEM may include regular meetings and virtual meetups with the community at large or with individuals, onboarding and offboarding of community members, community health checks, running workshops and other events, and actively building the community via networking, attending conferences, and working with other communities of practice. The final role, the RCM, is a blend of these two responsibilities and is likely to have technical understanding and the ability to engage with the community, but unlikely to be responsible for strategic decisions around tool design or running large workshops by themselves. RCMs also facilitate collaboration within the project team to ensure different skills are combined to respond to technical and non-technical needs in the project and its community. Teams of community managers can therefore consist of one or many of these three roles depending on the needs of the community and the purpose of the tool.

The specialisation of roles allows for clearer job titles and descriptions, thus allowing community managers to be appropriately recruited and recognised for their contributions to a project.

Specialisation also allows more flexibility in how roles are hired. For example, RAMs may be hired at the start of a project to work with software developers to identify end-user needs and stakeholder mapping so that front-end tools can be accessible to the community. On the other hand, CEMs may be most important once a tool has been published and there is a need to grow and engage the community.

## The Future of Open Source Community Management

Maintaining and sustaining open source software will always be challenging and an area of interest for many. Whilst people are still talking about a turnkey solution, this simply cannot exist as 'open source' does not refer to a single thing but a myriad of possibilities. It seems likely that community manager roles are here to stay as an important profession to help design and manage solutions for long-term sustainability. However, unless community managers are themselves valued and looked after, the communities they manage won't have a chance to succeed.

The status quo sees community managers as unicorns responsible for developing strategies and development plans for both the community *and* the codebase, whilst splitting their loyalties between their communities and their employers. Instead, we discuss greater clarity over job roles, scenarios for when different types of community managers should be considered and qualitative metrics such as CHAOSS to make visible *all* of the tasks that a community manager performs. With these developments, there is a much greater possibility of community managers being engaged and happy, and these feelings being propagated throughout the community they're in charge of.